\documentclass[preprint2]{aastex}
\usepackage{spr-astr-addons}
\usepackage{graphicx}
\begin{document}

\title{Gemini Observations of Disks and Jets in
Young Stellar Objects and in Active Galaxies}

\shorttitle{Jets \& Disks in YSOs \& AGN}        
\shortauthors{McGregor et al.}

\author{Peter McGregor, Michael Dopita Ralph Sutherland}
\affil{Research School of Astronomy and Astrophysics, ANU, \\ Cotter Road,  Weston Creek, ACT2611, Australia}
\author{Tracy Beck}
\affil{Gemini Observatory, 670 N. A'ohoku Place, Hilo, HI 96720, USA}
\author{Thaisa Storchi-Bergmann}
\affil{Instituto de Fisica, Universidade Federal do Rio Grande do Sul, Porto Alegre, RS 91501, Brasil}



\begin{abstract}
We present first results from the Near-infrared Integral Field Spectrograph (NIFS) located at Gemini North. For the active galaxies Cygnus A and Perseus A we observe rotationally-supported accretion disks and adduce the existence of massive central black holes and estimate their masses. In Cygnus A we also see remarkable high-excitation ionization cones dominated by photoionization from the central engine. In the T-Tauri stars HV Tau C and DG Tau we see highly-collimated bipolar outflows in the [\ion{Fe}{2}] $\lambda 1.644$ micron line, surrounded by a slower molecular bipolar outflow seen in the H$_2$ lines, in accordance with the model advocated by \cite{Pyo02}.
\end{abstract}



\section{The NIFS Instrument on Gemini North}\label{sec_1}
\begin{figure}
\begin{center}
  \includegraphics[width=0.45\textwidth]{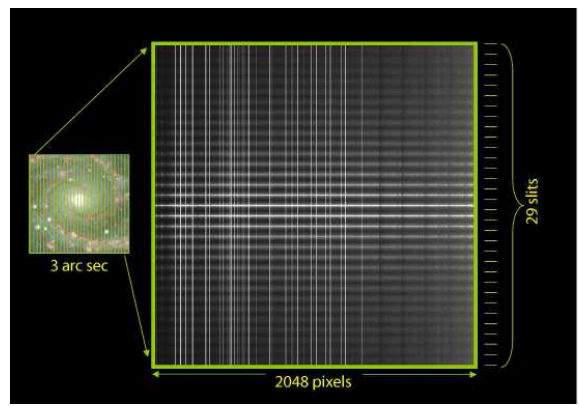}
 \end{center}    
\caption{The data format of NIFS, the concentric image slicer divides the field into 29 slitlets each $0.1''$ wide with $0.04''$ pixels in the spatial direction for a total length of $3.0"$ on the sky. Data reduction is therefore relatively simple, since the data effectively consists of 29 long-slit spectra.}
\label{fig_1}   
\end{figure}

The Research School of Astronomy and Astrophysics (RSAA) of the Australian National University (ANU) recently completed the Near-infrared Integral Field Spectrograph (NIFS) which is now located at Gemini North. A preliminary design for NIFS was described by \cite{McGregor99}, and the science mission by \cite{McGregor01}. 

The NIFS instrument utilizes an innovative concentric image slicing design and is designed to perform near diffraction-limited, near-infrared, imaging spectroscopy with the ALTAIR facility adaptive optics system on Gemini North. The image slicer divides the field into 29 slitlets each projecting to $0.1''$ on the sky and with $0.04''$ pixels in the spatial direction. It is designed to operate in each of the $J$, $H$, and $K$ photometric pass bands. NIFS is housed in a duplicate of the NIRI cryostat. It includes the Gemini On-Instrument Wavefront Sensor (OIWFS), mechanism and temperature control systems, and EPICS control software.

The detector is a $2048 \times 2048$ HgCdTe HAWAII-2 PACE technology array manufactured by the Rockwell Science Center, developed in collaboration with the Institute for Astronomy of the University of Hawaii. The data format given by the NIFS instrument is illustrated in Figure \ref{fig_1}

In summary the performance characteristics of the NIFS instrument are as follows:
\label{intro}
\begin{itemize}
\item{29 slitlets, each with 2040 spectral pixels, giving a total field of view of  $3.0 \times 3.0$~arcsec.}
\item{Pixel size of 0.04 x 0.10~arcsec.}
\item{A resolution of $R\sim 5300$, corresponding to a velocity resolution of $\sim$ 60 km~s$^{-1}$}
\item{Operation in the Z, J, H and K wavebands}
\item{Images corrected by adaptive optics (AO)}
\item{Image quality reaching down to the diffraction limit: 0.04~arcsec. in  J, 0.05~arcsec. in H, \& 0.06~arcsec.  in K}
\end{itemize}

In this paper, we describe some preliminary results obtained during both the commissioning time, and the guaranteed time on both pre-main sequence T-Tauri stars displaying both accretion disks and jets, and on two extragalactic sources, Cygnus A (3C~405, 4C~+40.40, IRAS~19577+4035) and Perseus A (NGC~1275, 3C~084, 4C~+41.07, IRAS~03164+4119), both of which show clear evidence of disks around a massive central nuclear black hole.

\section{Data Acquisition \& Reduction}\label{data}
The data described below were obtained on the Gemini North telescope in commissioning time and in guaranteed time between the dates of October 2005 and July 2006. The details will be given elsewhere. For these observations, the natural guide star system was used, with the nucleus (be it star or AGN) used for high order correction and a nearby bright offset guide star used as the on-instrument wavefront sensor (OIWFS) guide star in the cases where one existd. Typically, exposures ranged between 600 and 900~sec.  Equivalent length sky exposures were taken either side of these observations.

Data reduction was performed in the following order using IRAF scripts especially written for NIFS. Dark subtraction, reduction including flat fielding, using a bad pixel mask to fix hot pixels, transforming the image by stretching in $x$ and $y$ based on a Ronchi mask, sky subtraction (sky images were taken in nodding pattern), applying a telluric correction to both object and standard stars, then using these standard stars to flux calibrate using their magnitudes as given in the 2MASS catalog. An arc was taken immediately after the object and the wavelength calibration based on this arc had a rms fitting accuracy of 0.1~\AA. The final data cube obtained was in the form of a multi-extension FITS file with 2040 pixels in wavelength, and 29x69 spatial pixels.

\begin{figure}
\begin{center}
  \includegraphics[width=0.45\textwidth]{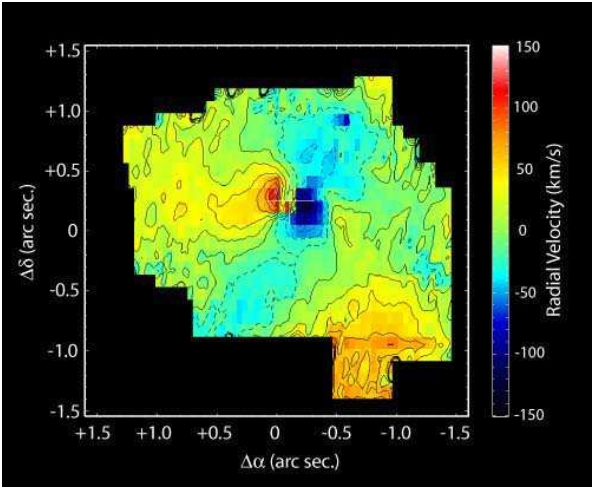}
 \end{center}    
\caption{The radial velocity field of the H$_2$ line in Perseus A (NGC~1275). This is consistent with Keplerian rotation of a warped accretion disk seen almost face on around a massive central black hole.}
\label{fig_2}   
\end{figure}

\begin{figure}
\begin{center}
  \includegraphics[width=0.45\textwidth]{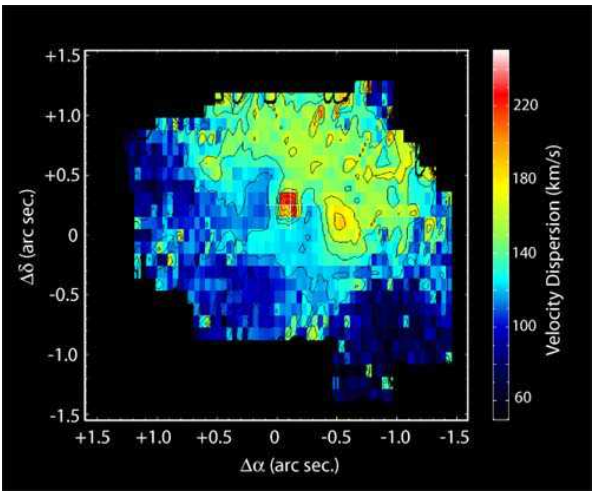}
 \end{center}    
\caption{The velocity dispersion of the H$_2$ line in Perseus A (NGC~1275). Note how it increases sharply towards the nucleus. This is consistent with a model in which the disk is predominately heated by turbulence and shocks as the molecular material works its way towards the nucleus. In this model the surface brightness of the disk is predicted to increase sharply towards the nucleus, in accord with the observational results.}
\label{fig_3}   
\end{figure}

\section{Active Galaxies}\label{AGN}
\subsection{Perseus~A}\label{Perseus}
The powerful radio galaxy Perseus A (NGC~1275), located at a redshift of 0.017559 (5264~km~s$^{-1}$), is the central component of the Perseus cluster, Abell~426. It is known by a number of other names of which the most frequently used are Perseus~A, 3C~84, 4C+41.07 and IRAS 03164+4119. The  cluster Abell~426 is composed of two distinct velocity components in the process of merging: a low velocity system moving at 5300~km~s$^{-1}$, which contains NGC~1275, and a high velocity component moving at 8200~km~s$^{-1}$ \citep{Rudy93,Krabbe00}. NGC~1275 is a star-forming early-type cD galaxy with an active nucleus. Over time, it has been assigned a whole host of classifications including peculiar Seyfert~1, Fanaroff-Riley~1 radio galaxy (F-R I), BL Lac object, a cooling flow galaxy, a LINER, and a post-merger galaxy.  

On the large scale, this galaxy is a strong source of thermal X-rays distributed with a complex morphology over a region of several arc minutes centred on the active nucleus \citep{Fabian00,Fabian03a}. It is also surrounded by a large and complex filamentary H$\alpha$ nebulosity \citep{Minkowski57,Lynds70} which has been recently studied in detail by \cite{Conselice01}. These filaments were believed to have been formed in a cooling flow, by the cooling of the intracluster medium \citep{Fabian94}. However, the complex bubble-like morphology of the X-ray gas and the intimate connection between X-ray and H$\alpha$ morphologies has led Fabian more recently to propose that these filaments are formed by the interaction of the radio plasma with the thermal gas in the galaxy \citep{Fabian03b}. This picture now is much more in accord with the standard picture of the interaction of strong radio jets with a surrounding medium, as detailed by \cite{Bicknell97} and \cite{Bicknell00}.

The small-scale dynamical structure was first studied by \citet{Wilman05}, who on the basis of long-slit spectra adduced the presence of a rotating disk of material in the nucleus, and who on this basis, estimated a black hole mass. However, both their spatial resolution and spatial coverage were inadequate for the purposes of establishing a reliable mass. 

The detailed disk structure as revealed by NIFS is remarkable, and is illustrated in Figures \ref{fig_2} and \ref{fig_3}. The inner portions of the molecular hydrogen and  [\ion{Fe}{2}] line images are best interpreted as a turbulent disk around the central active nucleus seen at a relatively small angle of inclination, and warped in its outer portions. The central inclination can be obtained from ellipse fitting of the central 0.5~arc sec of the H$_2$~1-0~S(1) summed line emission image, which yields $i=37.1\pm4.5^{\circ}$.

The surface brightness in the H$_2$~1-0~S(1) line falls off extremely steeply with radius, and fits well to a power-law in the range $0.2-1.2$ arc~sec., or radii in the range $40-400$pc. Such a surface brightness profile can be interpreted in the context of the model of \citet{Wilman05}, which postulates that the molecular hydrogen emission results from turbulent dissipation and shocks in an unstable or clumpy accretion disk.  In this, the turbulent dissipation of orbital energy results in an increase in binding energy of the gas, resulting in a net mass inflow of the disk material. In the model, gas is repeatedly shocked as it cascades towards the nucleus. Therefore, the net increase of binding energy of  the accretion disk as a function of radius, must be matched by radiative losses through turbulent losses and radial accretion shocks.

As will be shown elsewhere \citep{Dopita07}, the increase in the velocity dispersion of the gas towards the centre, seen in figure \ref{fig_3} and the observed surface brightness are consistent with this idea, provided that the gravitational zone of influence of the black hole is approximately 0.5 arc~sec. on the sky, or about 180~pc in radius. In this respect, the LINER nucleus of NGC~1275 is similar to the central disk around the central massive black hole of M87, studied in detail with HST by \citet{Dopita97}.

The radial velocity field can be used to constrain the enclosed mass within $\sim 35$~pc, assumed here to be mass of the central black hole. However, because of the small angle of inclination, the constraints are rather poor. From the data shown in \ref{fig_2}, \citet{Dopita07} derive a value of $\log[M_{\rm BH}] = 8.85 \pm 0.25$. This estimate is appreciably larger than that of \citet{Wilman05} ($\log[M_{\rm BH}] = 8.58 \pm 0.18$), but these authors lacked spatial coverage and were unaware of just how nearly face-on the nuclear disk in NGC~1275 actually is.

\subsection{Cygnus~A}\label{Cygnus}
Cygnus A is, in bolometric terms, the most powerful radio-emitting AGN in the local universe. Indeed, its power is great enough for it to be regarded as a radio quasar. In this sense, its proximity makes it a uniquely interesting object for detailed study.

On the large scale, Cygnus A displays a magnificent pair of radio lobes with intense hot-spots, and very finely-collimated radio jet in the inner regions. \citet{Wilson06} have investigated the properties of the limb-brightened, prolate spheroidal cavity in X-rays revealed by the Chandra Observatory. This cavity is filled with hotter shocked intracluster swept up by the expanding bubble of relativistic plasma associated with the radio jet. The total kinetic power of the expansion is found to be $1.2 \times 10^{46}$ ergs s$^{-1}$, consistent with the estimates of \citet{Kino05}.

The circum-nuclear region shows a complex dusty bipolar structure, whose axis is very well aligned with the large-scale radio jets. This has been studied at a resolution comparable to that attained by NIFS by \citet{Canalizo03}. Their 0.05 arc~sec. resolution Keck II adaptive optics near-infrared images clearly show an unresolved nucleus between two spectacular ionization/scattering cones. There is clear evidence of copious quantities of dust in these central region accompanied with heavy nuclear obscuration.

\cite{Tadhunter00} measured the (continuum) K-band polarization and discovered $> 28$\% polarization on the nucleus and some 25\% polarization in an extended linear structure aligned NNW-SSW.

\begin{figure}
\begin{center}
  \includegraphics[width=0.45\textwidth]{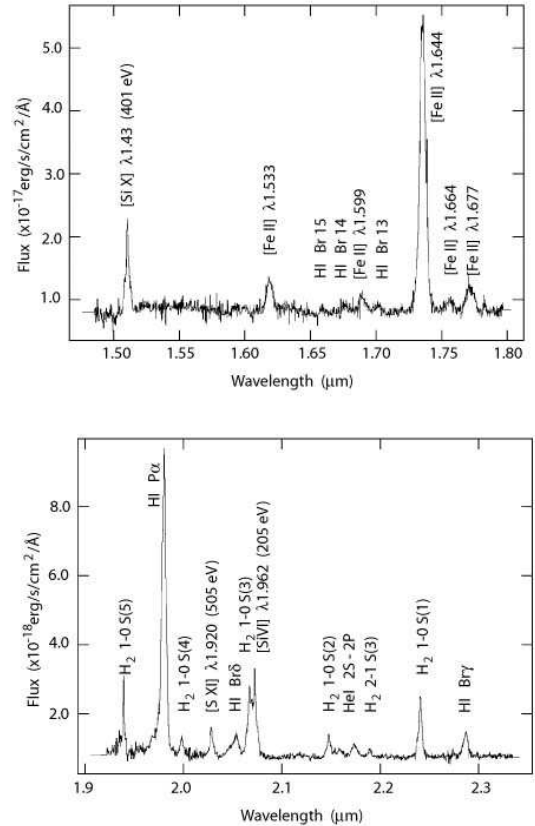}
 \end{center}    
\caption{The nuclear spectra of Cygnus A as observed in the H \& K bands. Note the very high signal to noise in these spectra. Note also the existence of molecular hydrogen, species of intermediate ionization and extremely highly ionized species such as [\ion{S}{11}] along this single line-of-sight. }
\label{fig_4}   
\end{figure}

IR spectroscopy of the nucleus in both the H \& K bands was obtained by \citet{Wilman00}, which revealed a rich emission line spectrum containing lines of molecular hydrogen, hydrogen and helium recombination lines, and forbidden lines of many species, some of which require photons of $>100$eV for their excitation. For comparison with these spectra, we present NIFS spectra of the nucleus in figure \ref{fig_4}. All of the lines identified by \citet{Wilman00} are clearly visible.

\begin{figure*}
\begin{center}
  \includegraphics[width=1.0\textwidth]{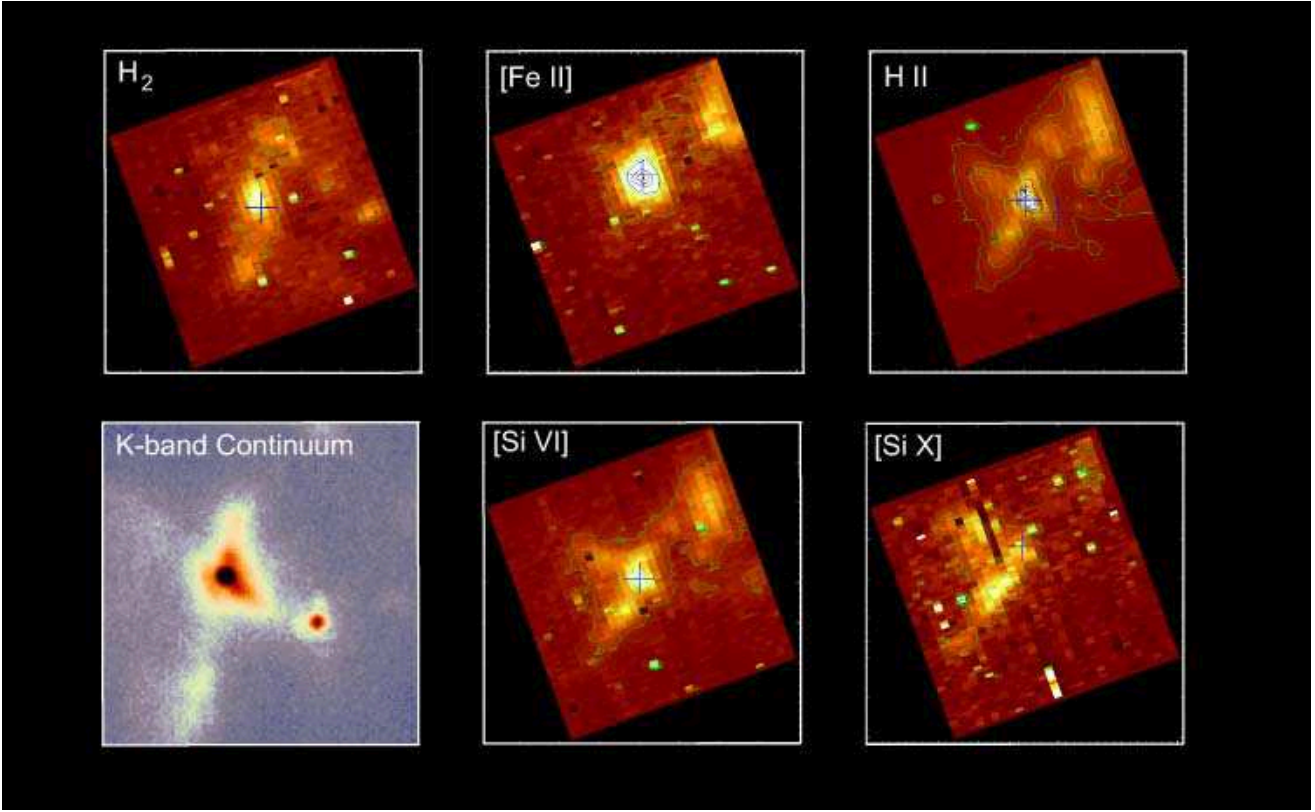}
 \end{center}    
\caption{The morphology of the nuclear regions of Cygnus A in various wavebands. North is at the top and east is at the left. The size of the white square is $4\times4"$. The NIFS line images are integrated over $\pm 500$km~s$^{-1}$ equivalent waveband. Note how the molecular hydrogen image reveals the bight inner accretion disk and a system of clouds which help define the ionization cone. The [\ion{Fe}{2}] line is confined to the circum-nuclear region and the innermost portions of the ionization cone, with a bright feature to the NW which likely represents a cloud illuminated by the central engine. Tthe remaining emission line images various aspects of the ionization cones discussed in the text.}
\label{fig_5}   
\end{figure*}

The spatial structure of the emission line regions observed with NIFS in the different ions is extremely instructive. This is shown for some of the brighter lines in figure \ref{fig_5}.

The broad-band K continuum \citep{Canalizo03} shows a classical biconical structure with a cone opening angle of about 70 degrees. The cones are orientated almost exactly along the direction of the radio jet. This morphology, and the fact that the westerly cone is brighter, suggests that we are seeing the inner region of a thick torus illuminated by the central  engine, seen nearly edge on, but inclined slightly such that the westerly radio jet is approaching the observer. There is also prominent point source to the SW of the nucleus which was suggested by \citet{Canalizo03} to be an extragalactic source. However, this source is not apparent on any of the NIFS images, and it represents something of a mystery. 

The structures in the ionized species highlight different aspects of the ionization cones. The easterly cone is most prominent in the most highly excited species. This suggests that this cone is characterized by a higher ionization parameter, $ \cal U$, than the westerly cone. It is possible that this cone is matter bounded, giving a high [\ion{Si}{6}] or [\ion{Si}{10}]  to \ion{H}{2} ratio. By contrast, in the W cone, dense matter seems to be obtruding into the cone, giving a large [\ion{Fe}{2}] to \ion{H}{2} ratio, but remaining relatively faint in the [\ion{Si}{10}] line emission. The  [\ion{Fe}{2}]  is also strong close to the nucleus, where it is presumably tracing the dense circum-nuclear gas in the inner torus.

The structure in the H$_2$ emission is remarkably different. On the large scale it shows an extended (ring or torus?) structure extending nearly N-S. Both the alignment and spatial extent of this structure match the the (continuum) K-band polarization structures traced by \cite{Tadhunter00}, which presumably arise from reflection of light from the central engine on the surface of dense and dusty clouds of gas. Separated from this by a low-level H$_2$ emission, there is a bright central region elongated at right angles to the jet direction. The inner region presumably traces the accretion disk proper, and the outer region a disk of material, possibly in the process of being accreted to the centre. This impression is confirmed by the detailed dynamics, which are consistent with a slowly-rotating outer region, and a much more rapidly rotating Keplerian inner disk warped with respect to the outer ring.

For Cygnus A, we may conclude that the outer structures help to define the directions of escape of the ionizing photons from the central engine, and thus the ionization cone structure seen in the highly ionized species. The ionization structure observed is consistent with photoionization in the radiative pressure dominated regime modelled by \citet{Groves04a,Groves04b}. It reaches very high ionization parameters in the easterly cone; ${\cal U} \sim 1.0$. Finally, we can tentatively conclude from the rapid rotation seen over the central $\sim 0.2$ arc~sec. that the enclosed mass within $\sim 100$ pc radius, assumed to be the mass of the central Black Hole, is of order $\log[M_{\rm BH}] = 9.5$.

\section{Young Stellar Objects (YSOs)}\label{YSOs}
\subsection{HV Tau C}\label{HV Tau}
The object HV Tau C is one component of a triplet of pre-main sequence stars \citep{Simon92}, and displays very strong emission lines \citep{Maguzzu94}. It  represents a fine example of an accretion disk around a YSO. From high resolution near infrared images, \citet{Monin00} infer that the disk is seen nearly edge-on ($i \sim 84^\circ$), and has a radius of $\sim 50$~AU. Very similar parameters were inferred from HST imaging by \cite{Stapelfeldt03}, $i \sim 84^\circ$ and $R \sim 80$~AU.  The disk is sufficiently inclined that (time-variable) H$_2$O ice absorption is seen in the IR with an optical depth in the circumstellar disk of order 1.5 mag. This implies A$_{\rm V} \sim 20$, \citep{Terada05}.

Spectroscopically, HV Tau C has been studied in the red region of the optical spectrum, and at high resolution by \cite{Apppenzeller05}. They found a slight asymmetry in the [\ion{N}{2}] lines, while the [\ion{S}{2}] $\lambda 6716$\AA line is practically symmetric. In the IR, \cite{Terada05} made a high-resolution $R = 10000$ study of the  [\ion{Fe}{2}] line at 1.644$\mu$m using the Subaru IRCS and adaptive optics with a 0.3 arc~sec. slit. Components at $\pm 35$km~s$^{-1}$ were seen, offset from one another by $\sim 0.5$ arc~sec., from which they inferred the existence of a jet with an opening angle of $\sim 13^\circ$.

\begin{figure}
\begin{center}
  \includegraphics[width=0.45\textwidth]{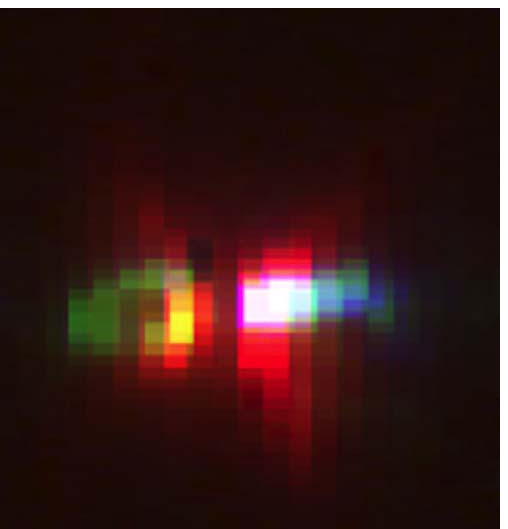}
 \end{center}    
\caption{The morphology of the jet and molecular outflow in HV Tau C. This image is 2 arc sec. on a side. North is at the top and east is at the left. The H$_2$~ $\lambda 2.122 \mu$m emission is shown in red. This has an outflow velocity of less than 50 km~s$^{-1}$, which is marginally detected in the NIFS data cube. The [\ion{Fe}{2}] line at 1.644$\mu$m is shown in the green and the blue channels. The green represents the velocity range -100 to -50  km~s$^{-1}$, while the blue represents the velocity range  -50 to 0 km~s$^{-1}$. Note the bi-conical H$_2$ region around the much faster expanding and well-collimated jet on the W side which has a full opening angle $< 30^\circ$. The eastern jet is apparently disrupted and forms a bubble-like structure.}
\label{fig_6}   
\end{figure}

The NIFS images stunningly reveal an  [\ion{Fe}{2}]  jet embedded in a slower H$_2$ conical bipolar outflow with a cone with total opening angle of about 120$^\circ$. This is shown in figure \ref{fig_6}. The  [\ion{Fe}{2}]  jet  is clearly visible in the -125 km~s$^{-1}$ velocity channel to the E, and at  +25 km~s$^{-1}$ to the W (velocities given are Heliocentric radial velocities). For an inclination angle of only 6$^\circ$, this would imply a peak outflow velocity of $\sim 700$km~s$^{-1}$. This is probably too high, and the true inclination angle is more like $\sim 10^\circ$, which would give a true outflow velocity of $\sim 400$km~s$^{-1}$.

This picture of a strongly-collimated fast jet, probably ionized by the working shocks within it, and surrounded by a slower, less collimated, molecular outflow is in good accord with the model advocated by \cite{Pyo02} for other T-Tauri outflows.

\subsection{DG Tau}\label{DG Tau}
DG Tau B is another bright T-Tau star with strong evidence for jet outflows. It was imaged using HST and NICMOS by \citet{Padgett99}. It shows a sharply defined conical reflection region extending out to 400 AU and with a full opening angle of 80$^\circ$ on the near side, and a less clearly delineated outflow on the far side. The absorbing equatorial disk is clearly seen.

The jet outflow was subjected to optical spectro-imaging observations at 0.5 arc~sec. resolution with OASIS  by \citet{Lavalley00}.This revealed a fast and unstable jet core surrounded by a slower moving flow. Line ratios are consistent with shocks with speeds of $50-100$ km s$^{-1}$ with increasing flow velocity, and decreasing density away from the star.

\citet{Takami04} studied the near-infrared H$_2$ emission in DG Tau using the Infrared Camera and Spectrograph (IRCS) on the 8.2-m SUBARU telescope and found evidence for a spatial extension at right angles to the jet axis. The line flux ratios of the 1-0 S(0) and 1-0 S(1) lines and an upper limit for the 2-1 S(1) to 1-0 S(1) H$_2$ line ratio suggests that the molecular  flow is thermalized at a temperature of $\sim 2000$ K, and is likely heated by shocks. 

 \citet{Pyo05} has studied the dynamics of DG Tau in the  [\ion{Fe}{2}] line at 1.644$\mu$m with an angular resolution of up to 0.16 arc~sec. achieved using the Adaptive Optics System of Subaru Telescope. They detected two distinct velocity components separated in space and velocity. The high velocity, spatially extended components show radial velocities $>250$ km s$^{-1}$ while the low velocity components lie close the the exciting star and have peak velocities in the range $80-150$ km s$^{-1}$.

\begin{figure}
\begin{center}
  \includegraphics[width=0.45\textwidth]{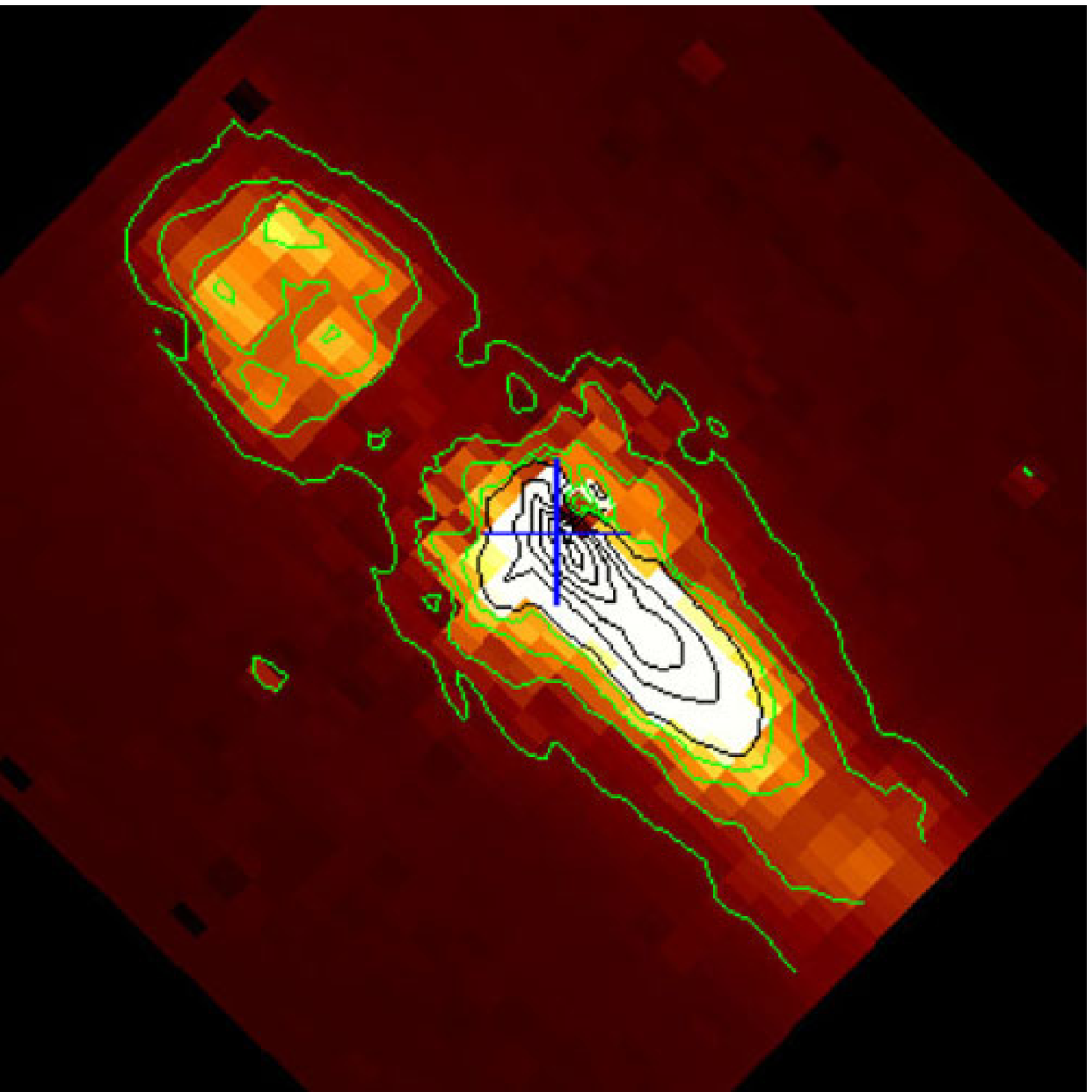}
 \end{center}    
\caption{The morphology of the DG Tau jet as seen by NIFS in the integrated light of the [\ion{Fe}{2}] line at 1.644$\mu$m. North is at the top east at the left, and the image is $3\times3$arc~sec in size, which corresponds to 460 AU. Note that the (inclined) equatorial disk is clearly visible, and is therefore still optically thick at 1.644$\mu$m within a projected distance of about 100 AU .}
\label{fig_7}   
\end{figure}

The NIFS image in the integrated light of the [\ion{Fe}{2}] line at 1.644$\mu$m is shown in figure \ref{fig_7}. The jet is very clearly defined on the approaching side, but like HV Tau C, it is less well defined and has a morphology more like an expanding bubble on its receding side. On the approaching side, the velocity channels show a sharply defined core of emission, showing slight wiggles in spatial coordinates and displaying a smooth velocity gradient on the approaching side, Its velocities range from -100 km s$^{-1}$ within 0.5 arc~sec. of the star, and up to -300 km s$^{-1}$  1.5 arc~sec. away from the star. At lower velocities, and also on the approaching side, a conical feature is seen with an opening angle of about 25$^\circ$. This too displays a smooth velocity gradient ranging from -75 km s$^{-1}$ within 0.5 arc~sec. of the star, up to -180 km~s$^{-1}$  1.5 arc~sec. away. Thus there is clear dynamical evidence for two distinct  [\ion{Fe}{2}] regions - a fast jet, and a slower sheath.

The receding lobe (where it is not obscured by the central equatorial disk) looks much more like an expanding bubble of gas, except for an axial enhancement in brightness seen at velocities +100 to +260 km s$^{-1}$, which probably represents shocks in the receding jet.

The smooth acceleration of both the jet and its sheath require that there be a source of energy in the outflowing gas, almost certainly stored magnetic energy. Thus, we agree with \citet{Lavalley00} that the jet must be dominated by its magnetic energy, as proposed in the X-wind model of \citet{Shu94}, the disk models of \cite{Kuker04} or the ambipolar diffusion model of \citet{Garcia01}. These classes of model are reviewed by \citet{GomezdeCastro04}. These magnetic jet models offer a natural way to explain the (otherwise puzzling) observations of a soft X-ray jet in DG Tau A by \citet{Gudel04}, which can be interpreted as resulting from thermal emission from shocks in the turbulent sheath of an accelerating magnetically-dominated jet.

We are undertaking hydrodynamical modelling of such jets using the new \emph{Fyris Alpha} code developed by Sutherland. This is a 3D multi-level refined mesh PPM code which uses a nested and adaptive grid with Lagrangian re-map. It includes both atomic and molecular radiation, both time and space variable equation of state and  ionisation, and self - gravity. The initial computational and physical parameters determined for the DG Tau jet system are as follows:
\begin{itemize}
\item{Computation box: $10^{16}$ cm on a side with 3 levels of resolution giving a maximal resolution of  $7.72\times10^{12}$ cm, or 0.51 AU.}
\item{ Gravity:   Fixed isothermal spherical self-gravity potential with $<V> = 4$ km/s  and $R_{\rm core} = 2.0\times10^{15}$ cm (or 133 AU). This potential puts a mass of $M = 2.81M_{\odot}$ within 100 AU.} 
\item{A three component disk \& disk atmosphere consisting of a  thin Keplerian disk $T< 30$K, a thick disk which is gravity and gas pressure supported with $T \sim 300$ ~K and a warm atomic atmosphere with  $T \sim 300$ ~K, all components being in intial pressure equilibrium.}
\item{ A ``heavyÓ  stellar jet with the following parameters: outflow velocity $v= 300$ km~s$^{-1}$, radius $R \sim 1 \times10^{14} $ cm and mass-loss rate $10^{-12} - 10^{-7} M_{\odot}$~yr$^{-1}$.}
\end{itemize}

These models establish the timescales of the problem, show that the observed morphologies can be reproduced in the models, and establish that there is an apparently smooth acceleration in the  entrained sheath of the jet. However, these models cannot yet be considered in any sense definitive.

Jet acceleration occurs naturally in any supersonic flow in a divergent cavity, in this case formed by the dense disk material. This effect could be enhanced by the presence of a strong internal magnetic field. This provides an additional energy source to the expanding flow generating magneto-hydrodynamic acceleration. Neither of these mechanisms have been tested rigourously in a self-consistent simulation, and there is little theoretical work on late-time jet acceleration, as distinct from acceleration mechanisms associated with the launch region.  Indeed even the launch mechanism is subject to considerable debate, although an MHD origin is the most likely hypothesis.

It remains possible that the apparently steady acceleration over the first $\sim 1.0$ arc~sec. of the jet flow is simply due to an observational selection effect. If the jet material was cool and intrinsically faint, only visible in small knots in the highest velocity slices, then the remaining emission could all be due to the shocked and entrained material that is progressively accelerated to higher speeds along the jet outflow. If this is the case then the NIFs observation may have resolved the entrainment and mass loading region of the jet, rather than the jet itself, or the actual launch region of the jet.

The timescales associated with significant changes in the jet are exceedingly short, and gross morphological changes are expected within a year or two. It is therefore very important to make time-resolved series of observations on such T-Tauri jets.

\begin{acknowledgements}
M.D. acknowledges the support of the ANU and the Australian Research Council (ARC) for his ARC Australian Federation Fellowship. The ARC also supports M.D. \& PMcG. under ARC Discovery Project DP0342844, and M.D. \& R.S. under  ARC Discovery Project DP0664434. 
\end{acknowledgements}




\end{document}